\documentclass[aps,prd,twocolumn,showpacs,groupedaddress,nofootinbib,amsmath]{revtex4}
\usepackage{graphicx}
\usepackage{epsfig}
\usepackage{multirow}
\usepackage{subfigure}
\newcommand{\be}{\begin{equation}}
\newcommand{\ee}{\end{equation}}
\newcommand{\ba}{\begin{eqnarray}}
\newcommand{\ea}{\end{eqnarray}}

\newcommand{\Dslash}{\lower-0.18ex\hbox{\makebox[-1pt][l]{\,/}}D}

\begin{document}

\title{Masses of the conjectured H-dibaryon for different channels at different temperatures  }

\author{Liang-Kai Wu}
\affiliation{School of Physics and Electronic Engineering, Jiangsu University, Zhenjiang, 212013, People's Republic of China}

\author{Xi-Rui Zhao}
\affiliation{School of Physics and Electronic Engineering, Jiangsu University, Zhenjiang, 212013, People's Republic of China}

\author{Ning Li}
\thanks{Corresponding author. Email address: lining@ujs.edu.cn}
\affiliation{School of Physics and Electronic Engineering, Jiangsu University, Zhenjiang, 212013, People's Republic of China}

\author{Yong-Liang Hao}
\thanks{Corresponding author. Email address: yhao@ujs.edu.cn}
\affiliation{School of Physics and Electronic Engineering, Jiangsu University, Zhenjiang, 212013, People's Republic of China}
\author{Xiao-Zhu Yu}
\thanks{Corresponding author. Email address: xzyu@ujs.edu.cn}
\affiliation{School of Physics and Electronic Engineering, Jiangsu University, Zhenjiang, 212013, People's Republic of China}

\date{\today}

\begin{abstract}
We present a lattice QCD spectroscopy study of
the conjectured H dibaryon for 5 different channels at nine different temperatures. The H dibaryon operator is constructed with five different channels which are flavor singlet, flavor 27-plet, $\Lambda \Lambda$, $N \Xi$ and $\Sigma \Sigma$.  The nine different temperatures range from
$T/T_c =0.24$ to $T/T_c = 1.90$.
The simulations are performed on anisotropic lattice with $N_f=2+1$ flavours of clover fermion  at quark mass
which corresponds to  $m_\pi=384(4) {\rm MeV} $.  The thermal ensembles were provided by the FASTSUM
collaboration and the zero temperature ensembles by the Hadspec collaboration.    The simulations show that the mass of H-dibaryon for 27-plet channel is the largest at different temperatures, while the mass for $\Sigma \Sigma $ channel is the lightest. We also calculate  the spectral function of the correlation function of H dibaryon for five channels.
The spectral density distributions exhibit similar behavior for the five channels.  The mass differences $\Delta m = m_H - 2\,m_{\Lambda} $ of H-dibaryon
 and
 $\Lambda$ pair  at $T/T_c =0.24 $  for five channels are also estimated. The results show that $\Delta m = m_H - 2\,m_{\Lambda} $ for channels of 27-plet and $\Lambda \Lambda$ is positive, while $\Delta m = m_H - 2\,m_{\Lambda} $ for channels of singlet, $N \Xi$ and $\Sigma \Sigma$ is negative.

\end{abstract}


\maketitle

\section{INTRODUCTION}
\label{SectionIntro}

 In addition to mesons and baryons, Quantum chromodynamics (QCD) predicts glueballs, hybrids, tetraquarks, pentaquarks and hexaquarks. 
  In 1976, by using the bag model, Jaffe predicted a flavour-singlet state ($uuddss$) with quantum number $I(J^P)=0(0^+) $ which is called
  H-dibaryon~\cite{Jaffe:1976yi}, and Jaffe predicted that the binding energy
   of H-dibaryon is about $80\ {\rm MeV}$ below the $\Lambda\Lambda$ threshold $2230\ {\rm MeV}$ which means that H-dibaryon is a deeply bound state.

   Unlike  mesons and baryons, this exotic hadron state may be of great importance  to the hypernuclei, and to the strange matter  which could exist
  in the core of neutron star.  Moreover, it is a potential candidate for dark matter~\cite{Farrar:2017eqq}. Theoretically, the physics of exotic states may
be different from those of the usual mesons and baryons.
Therefore,
  the prediction  by Jaffe triggered a vigorous search for such a state, both experimentally~\cite{Aoki:1991ip,KEK-PSE224:1998trj,Takahashi:2001nm,Yoon:2007aq,KEKE176:2009jzw,Nakazawa:2010zza,Belle:2013sba,BaBar:2018hpv,Ekawa:2018oqt}
  and theoretically~\cite{Iwasaki:1987db,Luo:2011ar,Luo:2007zzb,Mackenzie:1985vv,Pochinsky:1998zi,Wetzorke:1999rt,Wetzorke:2002mx,Beane:2009py,NPLQCD:2010ocs,
 Beane:2011zpa,NPLQCD:2011naw, NPLQCD:2012mex,Inoue:2010hs,Inoue:2010es,Inoue:2011ai,HALQCD:2019wsz,Sasaki:2016gpc,HALQCD:2018lur,Francis:2018qch,Green:2021qol}.

  Some experiments~\cite{Aoki:1991ip,KEK-PSE224:1998trj,Takahashi:2001nm,Yoon:2007aq,KEKE176:2009jzw,Nakazawa:2010zza,Ekawa:2018oqt}
  were carried out to search for H-dibaryon. Such search for H-dibaryon is nontrivial because of large mixing between hadronic states with various decay modes~\cite{Junnarkar:2024kwd}.
  The results from those experiments do not confirm the existence of H-dibaryon, but set the lower limit for the mass of H-dibaryon.

  The experiments~\cite{Belle:2013sba,BaBar:2018hpv} were carried out to search for H-dibaryon or deeply bound singlet $uuddss$ sexaquark  $S$
  (for the explanation of $S$,
  see~\cite{Farrar:2017eqq})
  in
  $\Upsilon \rightarrow S\bar\Lambda\bar\Lambda$ decay. Their results show no signal for the existence of H-dibaryon
  or $S$ particle.

   As a theoretically tool, Lattice QCD is also used to investigate H-dibaryon. Some quenched studies show that $m_H < 2\ m_{\Lambda}$,
   such results support that H-dibaryon is a bound state~\cite{Iwasaki:1987db,Luo:2011ar,Luo:2007zzb}
, while other quenched studies show that H-dibaryon is not a bound state~\cite{Mackenzie:1985vv,Pochinsky:1998zi,Wetzorke:1999rt,Wetzorke:2002mx}.

   Aside from quenched studies, simulations with dynamical fermions have been carried out.  Two methods  have been proposed.
   One is based on the L{\"u}scher's finite volume method~\cite{Luscher:1986pf,Luscher:1990ux}.  This method considers scattering between two baryons and extracts the scattering phase shift from the temporal behaviour of the hadronic correlations.  From the phase shift, whether the states are scattering states or binding states can be distinguished. Some work in this direction includes Refs.~\cite{Beane:2003da,Beane:2006mx,Beane:2009py,NPLQCD:2010ocs,Beane:2011zpa,NPLQCD:2011naw,NPLQCD:2012mex} by the NPLQCD collaboration and  Refs.~\cite{Francis:2018qch,Green:2021qol}.

 The other method focuses on the Nambu-Bethe-Salpeter(NBS)
 wave-function by computing the four-point green function on lattice, and then determined the baryon-baryon potential from the NBS
 wave-function.  The HALQCD collaboration  used this method to address the existence of H-dibaryon~\cite{Inoue:2010hs,Inoue:2010es,Inoue:2011ai,HALQCD:2019wsz,Sasaki:2016gpc,HALQCD:2018lur}.

Apart from the search for H-dibaryon, there are lattice QCD calculations to addressing dibaryon states with heavy  quark flavour (charm ($c$) or bottom ($b$) quark). Some work includes Refs.~\cite{Junnarkar:2024kwd,Lyu:2022tsd,Xing:2025uai,Junnarkar:2022yak,Mathur:2022ovu}.

Besides at zero temperature, the properties of hadrons at finite temperature are also one of the central goals of lattice QCD simulation
(see, for example,~\cite{Aarts:2010ek,Aarts:2011sm,Aarts:2013kaa,Aarts:2014cda,Kelly:2018hsi,Aarts:2020vyb}). In the past decades, mesons
at finite temperature have been studied
extensively~\cite{Karsch:2003jg}. However, this is not the case for baryons. Baryons at finite temperature are hardly investigated on the lattice.
In fact, there are a few lattice studies of baryonic screening and temporal masses \cite{DeTar:1987ar,Pushkina:2004wa,Aarts:2018glk,Datta:2012fz,Aarts:2015mma}
. Nevertheless, the behaviour of baryons and exotic states in a hadronic medium
is relevant to heavy-ion collisions. Therefore, there is a need to unambiguously understand the property  of baryons and exotic states at finite temperature.

Ref.~\cite{Wu:2023zmh} investigated the spectrum of H-dibaryon at finite temperature where the operator of H-dibaryon is made up of flavor singlet. The H-dibaryon can be
formulated with other channels which are flavor 27-plet,  $\Lambda \Lambda$, $N \Xi$ and $\Sigma \Sigma$~\cite{Francis:2018qch}.  Based on the simulations in Ref.~\cite{Wu:2023zmh},
 we make lattice QCD simulations to investigate the masses of the conjectured H-dibaryon with the other channels at different temperatures with
the goal
 to  gain some insights into the properties of H-dibaryon such as mass variation with temperature,  and its existence.

The paper is organized as follows: In Sec.~\ref{SectionLattice},
we present the  the definition of  the interpolating operators for the five channels of H-dibaryon and describe the extrapolation method.
  Our simulation results are
given in Sec.~\ref{SectionMC},  followed by discussion in
Sec.~\ref{SectionDiscussion}. The correlation function expressions are included in an appendix. 

\begin{figure}[h]
\centerline{\epsfxsize=3.0in\hspace*{0cm}\epsfbox{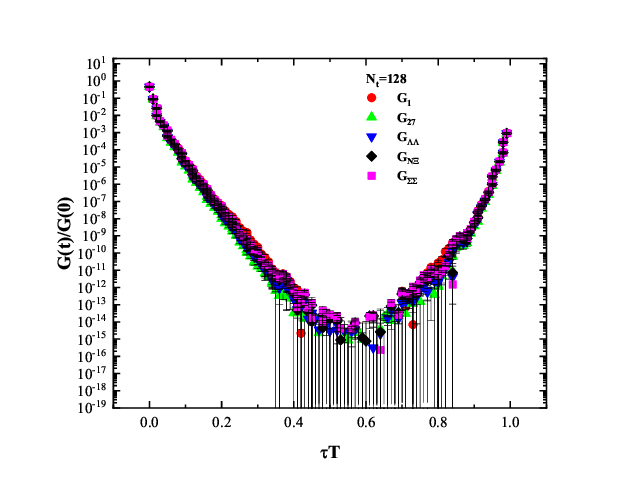}}%
\caption{\label{fig1} Euclidean correlators $G(\tau)/G(0) $ of $H_1$, $H_{27}$, $H_{\Lambda\Lambda}$, $H_{N\Xi}$ and $H_{\Sigma\Sigma}$ as a function of $\tau T$ at $N_t = 128$ lattice corresponding to the lowest temperature $T/T_c =0.24$.  The correlators at some points are not displayed due to the minus value.  The vertical axis is rescaled logarithmically}
\end{figure}

\begin{figure}[h]
\centerline{\epsfxsize=3.0in\hspace*{0cm}\epsfbox{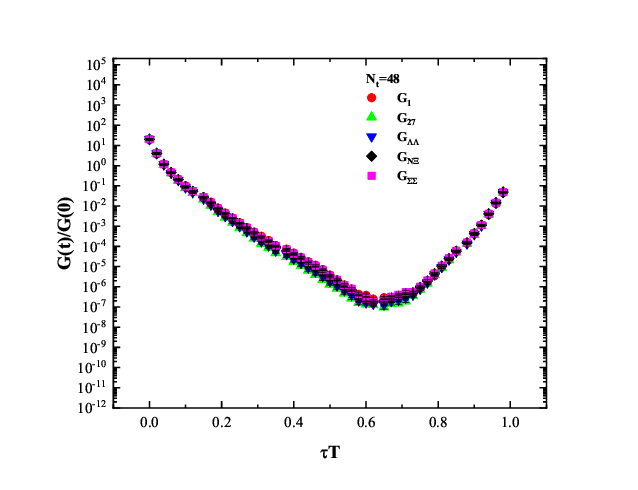}}%
\caption{\label{fig2} Euclidean correlator $G(\tau)/G(0) $ of $H_1$, $H_{27}$, $H_{\Lambda\Lambda}$, $H_{N\Xi}$ and $H_{\Sigma\Sigma}$ as a function of $\tau T$ at $N_t = 48$ lattice corresponding to the temperature $T/T_c =0.63$. The vertical axis is rescaled logarithmically }
\end{figure}

\begin{figure}[h]
\centerline{\epsfxsize=3.0in\hspace*{0cm}\epsfbox{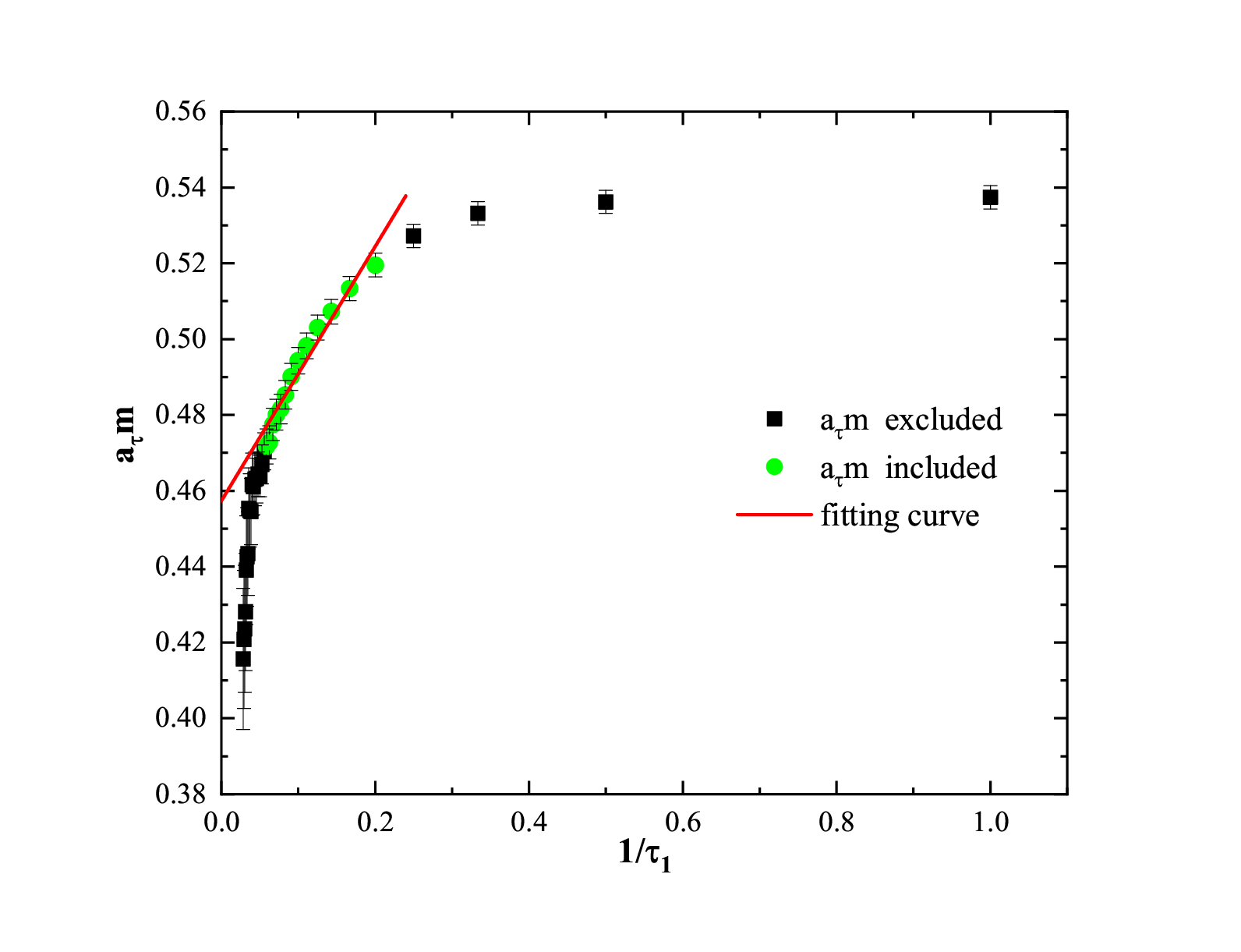}}%
\caption{\label{fitting_mass} Linear  extrapolation of mass values for $H_1$ on $N_\tau=128 $ ensembles. The data points indicated by solid squares are excluded in the extrapolation procedure, while the data points represented by solid circles are included. Horizontal axis represents inverse values of time slices suppressed. }
\end{figure}

\begin{figure}[h]
\centerline{\epsfxsize=3.0in\hspace*{0cm}\epsfbox{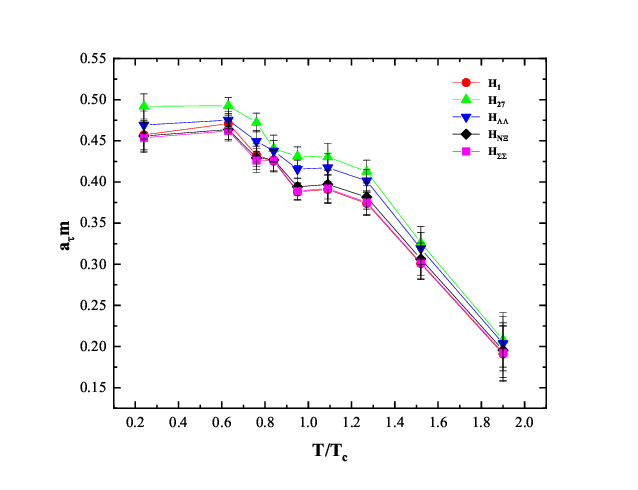}}%
\caption{\label{mass_vs_t} Mass values for $H_1$, $H_{27}$, $H_{\Lambda\Lambda}$, $H_{N\Xi}$ and $H_{\Sigma\Sigma}$ at different temperatures.  }
\end{figure}

\begin{figure}[h]
\centerline{\epsfxsize=3.0in\hspace*{0cm}\epsfbox{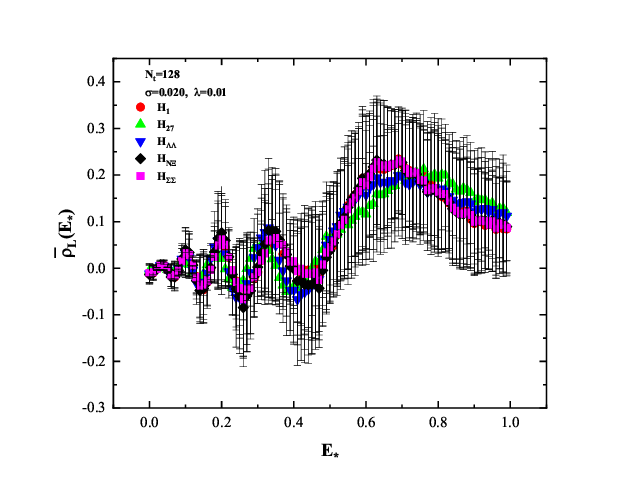}}%
\centerline{\epsfxsize=3.0in\hspace*{0cm}\epsfbox{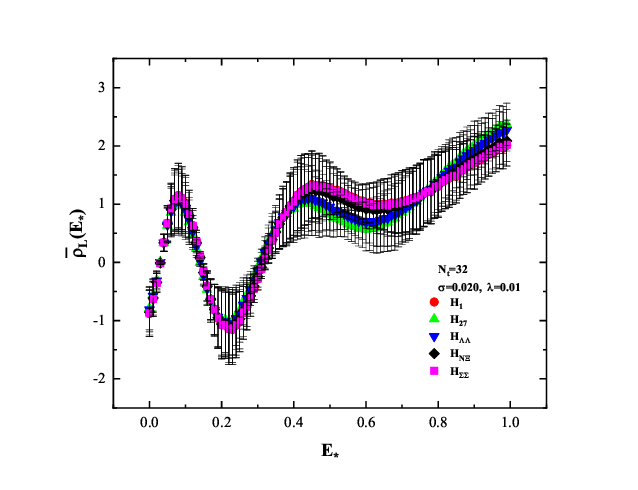}}
\centerline{\epsfxsize=3.0in\hspace*{0cm}\epsfbox{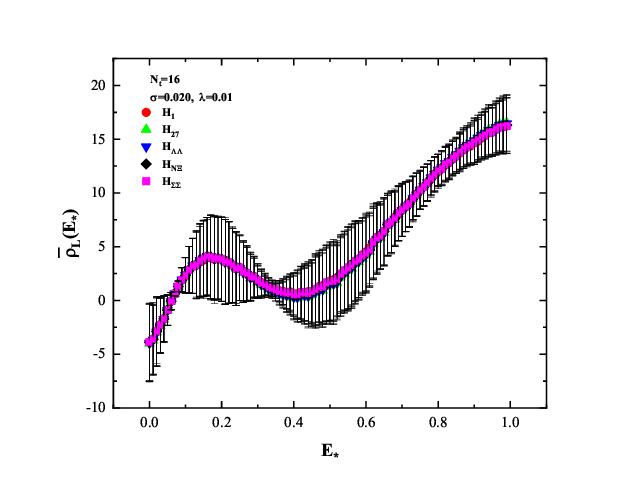}}
\caption{\label{Xi_spectral} Spectral density of $H_1$, $H_{27}$, $H_{\Lambda\Lambda}$, $H_{N\Xi}$ and $H_{\Sigma\Sigma}$ at different temperatures.  }
\end{figure}

\section{LATTICE CALCULATION AND SETUP}
\label{SectionLattice}
In our simulation, we compute the correlation functions of H-dibaryon for five channels which are flavor singlet, flavor 27-plet, $\Lambda \Lambda$, $N \Xi$ and $\Sigma \Sigma$~\cite{Francis:2018qch}. The five channels of H-dibaryon are represented by $H_1$, $H_{27}$, $H_{\Lambda\Lambda}$, $H_{N\Xi}$ and $H_{\Sigma\Sigma}$, respectively.

 The generic form of correlation function is:
\begin{eqnarray}
        \label{correlation_function}
          G(\vec{x},\tau) &= < O(\vec{x},\tau)O^+(0)>.
       \end{eqnarray}
       Before defining the H-didaryon interpolating operator, we introduce objects for three quark  and six quark combination.
       The object for six quark combination can be given by~\cite{Francis:2018qch}:
\begin{align}\label{eq:hexaquark}
[abcdef] = \ & \epsilon_{ijk} \epsilon_{lmn} \Big( b^i C\gamma_5 P_+ c^j
\Big)  \nonumber \\
& \times\Big( e^l C\gamma_5 P_+ f^m \Big) \Big( a^k C\gamma_5 P_+ d^n \Big) ({\vec{x}}, t)\,.
\end{align}
The three quark combination is expressed as~\cite{Francis:2018qch}:
\begin{align}
B_\alpha = [abc]_\alpha=\epsilon_{ijk}(b^i C \gamma_5 P_+ c^j)a^k_\alpha,
\end{align}
where $a, b,\ldots,f$ denote generic quark flavors, $\alpha$ represents dirac index and
$P_+=(1+\gamma_0)/2$  projects the quark fields to positive parity.

Based on the three quark combination, we can define:
\begin{align}
(BB) = B_\alpha (C \gamma_5 P_+ )_{\alpha \beta} B_\beta.
\end{align}

The H-didaryon interpolating operators for five channels are given by the following expression~\cite{Francis:2018qch,Donoghue:1986zd,Golowich:1992zw,Wetzorke:1999rt,Wetzorke:2001tgi}:
$H_1$, $H_{27}$, $H_{\Lambda\Lambda}$, $H_{N\Xi}$ and $H_{\Sigma\Sigma}$
\begin{align}
H_\mathbf{1} &= \frac{1}{48}\Big( [sudsud] - [udusds] - [dudsus]  \Big),
\end{align}
\begin{align}
H_\mathbf{27} &= \frac{1}{48\sqrt{3}}\Big( 3 [sudsud] + [udusds] + [dudsus]  \Big),
\end{align}
\begin{align}
H_{\Lambda\Lambda} &= \frac{1}{12} [sud][sud],
\end{align}
\begin{align}
H_{N\Xi} =&\frac{1}{36}\Big(  [uud][ssd] -[dud][ssu] \nonumber  \\ 
  & + [ssd][uud]  - [ssu][dud] \Big ),
\end{align}
\begin{align}
H_{\Sigma\Sigma} &= \frac{1}{36\sqrt{3}}\Big( 2[uus][dds] - [dus][uds] - [dus][dus] \nonumber \\
& -[ uds][dus] - [uds][uds] + 2 [uus][dds]  \Big).
\end{align}

The H-dibaryon correlation function $G_1$ for $H_1$  channel can be found in Ref.~\cite{Wetzorke:2001tgi}.
the other correlation functions $G_{27}$, $G_{\Lambda\Lambda}$, $G_{N\Xi}$ and $G_{\Sigma\Sigma}$ for $H_{27}$, $H_{\Lambda\Lambda}$, $H_{N\Xi}$ and $H_{\Sigma\Sigma}$ channels are collected in the appendix~\ref{appendix:a}. 

After we get the correlation function, the mass can be obtained by fitting the exponential ansatz:
\begin{align}
\label{eq:Ansatz}
G(\tau) =  A_{+} e^{-m_{+}\tau} +  A_{-} e^{-m_{-}(1/T-\tau)},
\end{align}
with  $m_{+}$ being the mass of the particle of interest, and $\tau$ in the interval $0 \leq \tau<1/T$ on a lattice at finite temperature $T$.

In order to get the ground state energy
in our procedure to get the mass, we take an extrapolation method.  We fit the Eq.~(\ref{eq:Ansatz}) to correlators
for a time range $[\tau_1,\tau_2]$ where $\tau_2$ is fixed to the  whole time extent,  and $\tau_1$ runs over several values from  $\tau_1=1, 2, 3, 4,...$, then we can get a series of mass values
one of which corresponds to different early Euclidean time slices suppression.
After we obtain the series of mass values against different $1/\tau_1$,  we can  extrapolate the series of  masses to $\tau_1 \to \infty$. We illustrate this procedure in 
Fig.~\ref{fitting_mass}.

Hadrons properties are encoded in spectral functions.  We also compute the spectral functions for $H_1$, $H_{27}$, $H_{\Lambda\Lambda}$, $H_{N\Xi}$ and $H_{\Sigma\Sigma}$
using the method designed in Ref.~\cite{Hansen:2019idp}.  Ref.~\cite{Wu:2023zmh} presented a concise description for the method which cites Ref.~\cite{Hansen:2019idp}.
More details are given in Ref.~\cite{Hansen:2019idp}.


\section{MC SIMULATION RESULTS}
\label{SectionMC}

The computation details are the same as those in Ref.~\cite{Wu:2023zmh,Aarts:2020vyb}. To render this paper self-contained, we briefly describe the 
computation details. The simulations are carried out on  $N_f=2+1$ Generation2 (Gen2)
FASTSUM ensembles~\cite{Aarts:2020vyb} of which the ensembles at the lowest temperature are provided by the HadSpec
collaboration~\cite{Edwards:2008ja,HadronSpectrum:2008xlg}. The ensembles are generated with a Symanzik-improved gauge action and a
   tadpole-improved clover fermion action, with stout-smeared links.    
   
   The parameters in the lattice action are recompiled in Table \ref{tab:parameters}. 
 The $N_f=2+1$ Gen2 ensembles correspond to a physical strange quark mass and a bare light quark mass of $a_\tau m_l=-0.0840$, yielding a pion mass
   of $m_\pi=384(4)$ MeV (see Table \ref{tab:lattice_spacings}). The corresponding physical parameters such as the lattice spacing, and the pion mass etc are collected in Table ~\ref{tab:lattice_spacings}. The computation details are presented in the following three tables~\ref{tab:parameters},
~\ref{tab:lattice_spacings}, and~\ref{tab:mass}~\cite{Aarts:2020vyb}. 

\begin{table*}[t]
\caption{Parameters in the lattice action. This table is recompiled from Ref.~\cite{Aarts:2020vyb}.}
\label{tab:parameters}
\begin{ruledtabular}
\centering
\begin{tabular}{ l  l }
gauge coupling (fixed-scale approach)     & $\beta = 1.5$ \\
tree-level  coefficients 					& $c_0=5/3,\,c_1=-1/12$ \\
bare gauge, fermion anisotropy 			& $\gamma_g = 4.3$, $\gamma_f = 3.399$ \\
ratio of bare anisotropies          &  $\nu = \gamma_g / \gamma_f = 1.265$ \\
spatial tadpole (without, with smeared links) & $u_s = 0.733566$, $\tilde{u}_s = 0.92674$ \;  \\
temporal tadpole (without, with smeared links) \; & $u_\tau = 1$, $\tilde u_\tau = 1$  \\
spatial, temporal clover coefficient 			& $c_s = 1.5893$, $c_\tau = 0.90278$ \\
stout smearing for spatial links 			& $\rho = 0.14$, isotropic, 2 steps \\
bare light quark mass for Gen2 		& $\hat m_{0, \rm light} = -0.0840$ \\
bare strange quark mass          		& $\hat m_{0, \rm strange} = -0.0743$ \\
light quark hopping parameter for Gen2 &  $\kappa_{\rm light} = 0.2780$ \\
strange quark hopping parameter              &  $\kappa_{\rm strange} = 0.2765$ \\
\end{tabular}
\end{ruledtabular}
\end{table*}

\begin{table}[t]
\caption{Parameters such as lattice spacing, pion mass etc collected from Ref.~\cite{Aarts:2020vyb} for Generation 2 ensemble.}
\begin{ruledtabular}
  \begin{center}
    \begin{tabular}[t]{cc}
	$a_\tau$ [fm] & 0.0350(2)   \\
	\; $a_\tau^{-1}$ [GeV] \; & 5.63(4)  \\
	$\xi=a_s/a_\tau$ &  3.444(6)    \\
	$a_s$ [fm] & \; 0.1205(8) \; \;  \\
	$N_s$   &   24   \\
	$m_\pi$ [MeV] & 384(4)  \\
	$m_\pi L$ & 5.63  \\
  \end{tabular}
  \end{center}
  \label{tab:lattice_spacings}
  \end{ruledtabular}
\end{table}

\begin{table*}[htp]
\caption{Spatial and temporal extent, temperature in MeV, number of configurations,  mass of $H_1$, $H_{27}$, $H_{\Lambda\Lambda}$, $H_{N\Xi}$ and $H_{\Sigma\Sigma}$.
 Estimates of statistical and systematic errors are
contained in the first and second brackets, respectively. The ensembles at the lowest temperatures were provided by HadSpec~\cite{Edwards:2008ja,HadronSpectrum:2008xlg} (Gen2).  }
\begin{ruledtabular}
  \begin{center}
     \begin{tabular}[t]{ccccclllll}
	$N_s $ & \; $N_\tau$ & \; $T {\rm  [MeV]} $  & \;   $T/T_c$  & \;  $N_{\rm cfg}$  & \; $a_\tau m_{H_1}$ & \;  $a_\tau m_{H_{27}}$ & \;  $ a_\tau m_{H_{\Lambda\Lambda}} $ & \;  $a_\tau m_{H_{N\Xi}}$ & \; $a_\tau m_{H_{\Sigma\Sigma}} $  \\
 	\hline
24 & 128 & 44 & 0.24 & 305  & 0.457 (17) (3) & 0.492 (15) (3) & 0.469 (17) (3) & 0.456 (18) (3) & 0.454 (18) (3)\\
32 & 48 & 117 & 0.63 & 601  & 0.471 (10) (4) & 0.493 (10) (4) & 0.475 (11) (4) & 0.464 (12) (4) & 0.462 (12) (3)\\
24 & 40 & 141 & 0.76 & 502  & 0.433 (14) (14) & 0.472 (11) (15) & 0.450 (13) (16) & 0.429 (14) (15) & 0.426 (15) (14)\\
24 & 36 & 156 & 0.84 & 501  & 0.425 (13) (15) & 0.440 (16) (19) & 0.438 (13) (17) & 0.427 (13) (16) & 0.426 (13) (15)\\
24 & 32 & 176 & 0.95 & 1000  & 0.388 (10) (17) & 0.431 (11) (19) & 0.416 (11) (19) & 0.394 (11) (18) & 0.389 (10) (17)\\
24 & 28 & 201 & 1.09 & 1001  & 0.391 (17) (15) & 0.430 (16) (18) & 0.417 (17) (17) & 0.397 (18) (16) & 0.392 (17) (15)\\
24 & 24 & 235 & 1.27 & 1002  & 0.374 (15) (34) & 0.413 (14) (34) & 0.401 (14) (34) & 0.382 (15) (35) & 0.375 (15) (34)\\
24 & 20 & 281 & 1.52 & 1000  & 0.30 (2) (5) & 0.33 (2) (5) & 0.32 (2) (5) & 0.31 (2) (5) & 0.30 (2) (5)\\
24 & 16 & 352 & 1.90 & 1000  & 0.19 (3) (4) & 0.21 (3) (5) & 0.20 (3) (5) & 0.20 (3) (5) & 0.19 (3) (5)\\

    \end{tabular}
    \end{center}

    \label{tab:mass}
    \end{ruledtabular}
\end{table*}




The quark propagators are computed by using the deflation-accelerated algorithm~\cite{Luscher:2007es,Luscher:2007se}.  When computing the propagator,
The spatial links are stout smeared~\cite{Morningstar:2003gk} with two steps of smearing, using the weight $\rho = 0.14$. For the sources and sinks, we use the Gaussian smearing~\cite{Gusken:1989ad}
\begin{align}
\eta' = C\left(1+\kappa H\right)^n\eta,
\end{align}
where $H$ is the spatial hopping part of the Dirac operator and $C$ an appropriate normalisation~\cite{Aarts:2015mma}.

The correlators of $H_1$, $H_{27}$, $H_{\Lambda\Lambda}$, $H_{N\Xi}$ and $H_{\Sigma\Sigma}$ on lattice $N_t=128$ are presented in Fig.~\ref{fig1}, while the correlators on lattice $N_t =48 $ are plotted in Fig.~\ref{fig2}. Some correlator data points are negative on lattice $N_t=128$, and these points are not displayed on the plot, because the vertical axis is rescaled logarithmically. From Fig.~\ref{fig1} and Fig.~\ref{fig2}, we can find that the correlator behaviour of $H_1$, $H_{27}$, $H_{\Lambda\Lambda}$, $H_{N\Xi}$ and $H_{\Sigma\Sigma}$ is almost the same on the same lattice. 
At some points, the error bars look strange, it is because at these points, the errors are the magnitude of the correlator values, and the
 vertical axis is rescaled.  
 The correlators of $H_1$, $H_{27}$, $H_{\Lambda\Lambda}$, $H_{N\Xi}$ and $H_{\Sigma\Sigma}$ on other lattices are similar as those on lattice $N_t =48 $. 
 
 We use the extrapolation method to extract the ground state masses for  $H_1$, $H_{27}$, $H_{\Lambda\Lambda}$, $H_{N\Xi}$ and $H_{\Sigma\Sigma}$.  It is because the excited states have more effects at the early
 Euclidean time~\cite{Aarts:2015mma}, so we first fit equation~(\ref{eq:Ansatz}) to correlator
by suppressing different early time slices  to get a series of mass values. 
After we get a series of mass values with different early time slices suppressed, we  extrapolate the mass values linearly with the scenario described in the last paragraph in Sec.~\ref{SectionLattice}. We present the results of linear extrapolation for $H_1$ on lattice $N_\tau=128$ in Fig.~\ref{fitting_mass} where  the data points indicated by solid squares are excluded in the extrapolation procedure, while the data points represented by solid circles are included. 

The ground state masses for  $H_1$, $H_{27}$, $H_{\Lambda\Lambda}$, $H_{N\Xi}$ and $H_{\Sigma\Sigma}$ are listed in table~\ref{tab:mass}. We also plot the mass values of $H_1$, $H_{27}$, $H_{\Lambda\Lambda}$, $H_{N\Xi}$ and $H_{\Sigma\Sigma}$ in Fig.~\ref{mass_vs_t} from which
we can find that the masses decrease when temperature increases.  From table~\ref{tab:mass}, we can find that among the five channels for H-dibaryon,  the mass of 27-plet channel is the largest, while the mass of $\Sigma \Sigma $ channel is the lightest at different temperatures.

The mass values for $H_1$ are slightly different with those in Ref.~\cite{Wu:2023zmh}. It is because in Ref.~\cite{Wu:2023zmh}, the correlator of every configuration is rescaled relative to the first time slice value of their own. However, in this paper, the correlators of all configuration are rescaled with respect to the first time slice value of the first configuration.

 We also calculate the spectral density $\bar\rho_L(E_\star)$  of the correlation function of $H_1$, $H_{27}$, $H_{\Lambda\Lambda}$, $H_{N\Xi}$ and $H_{\Sigma\Sigma}$  by using the public computer program~\cite{Hansen:code}. In Ref.~\cite{Wu:2023zmh}, it is pointed out that among the $\sigma$ values $\sigma=0.02, 0.04, 0.06, 0.08$,  $\sigma =0.020$  is suitable for spectral density extraction from correlators. At low temperature, smaller $\sigma $ value  manifests rich peak structure of
spectral density in small $E_\star$ region. At intermediate temperature,  smaller $\sigma $ value can make  peak structure of
spectral density in small $E_\star $ region more pronounced. At high temperature, different $\sigma $ value has little difference on the computation of
spectral density. Therefore, we just present the results of spectral density for $H_1$, $H_{27}$, $H_{\Lambda\Lambda}$, $H_{N\Xi}$ and $H_{\Sigma\Sigma}$ computed at $\sigma =0.020$ in Fig.~\ref{Xi_spectral}.

Fig.~\ref{Xi_spectral} presents
the spectral density $\bar\rho_L(E_\star)$ of $H_1$, $H_{27}$, $H_{\Lambda\Lambda}$, $H_{N\Xi}$ and $H_{\Sigma\Sigma}$ for three temperatures.  The upper panel gives the spectral density $\bar\rho_L(E_\star)$ of $H_1$, $H_{27}$, $H_{\Lambda\Lambda}$, $H_{N\Xi}$ and $H_{\Sigma\Sigma}$ on lattice $N_t=128 $ which corresponds to temperature $T/T_c=0.24$.  The middle panel shows the results of $\bar\rho_L(E_\star)$ on lattice $N_t=32$ which corresponds to temperature $T/T_c=0.95$, while the lower panel  represents $\bar\rho_L(E_\star)$ on $N_t=16$ which corresponds to temperature $T/T_c=1.90$.  
 
 From Fig.~\ref{Xi_spectral}, one can find that at three temperatures, spectral density $\bar\rho_L(E_\star)$ of five channels of H-dibaryon has similar structure. At the lowest temperature, spectral density $\bar\rho_L(E_\star)$ of five channels of H-dibaryon has a rich peak structure.  The mass values of five channels which are approximately  $am \approx 0.44 $ at $T/T_c = 0.24$  are in the neighbour of peak position $E_\star = 0.35$,  reflecting more influences of excited states.
  At the intermediate temperature, the rich peak structure turns to a two-peak structure.   The
 mass values of five channels of H-dibaryon  presented in Table.~\ref{tab:mass} are not consistent with the peak positions of corresponding spectral density. This observation shows that the mass values obtained
 by fitting procedure are affected by the two-peak structure of spectral density.
 At the highest 
 temperature,  one peak can be found from Fig.~\ref{Xi_spectral}. The peak position is consistent with the mass value  obtained
 by fitting procedure.

\section{DISCUSSIONS}\label{SectionDiscussion}

We have made a simulation in an attempt to determine the masses of five channels of the conjectured particle H-dibaryon with $2+1$ flavor QCD with clover fermion
at nine different temperatures.
 The results  are collected in table~\ref{tab:mass}.  The
spectral density distribution of those particle's correlation function are computed to understand the mass spectrum obtained by fitting procedure.

From the simulation, one can find that the logarithmically rescaled correlators  of the five channels have little difference,  and the spectral density of the five channels shows almost the same structure. However, using the extrapolation method described in the last paragraph in Sec.~\ref{SectionLattice}, one can obtain the mass values of the five channels presented in table~\ref{tab:mass}. 

At the lowest temperature, the $a_\tau m_{H_{27}} = 0.492(15)(3)$ of the $H_{27}$ is the largest, while $a_\tau m_{H_{1}}=0.457(17)(3), a_\tau m_{H_{\Lambda\Lambda}}=0.456(18)(3), a_\tau m_{H_{\Sigma\Sigma}}=0.454(18)(3)$  have little difference.  Using the mass value of $\Lambda$ baryon in table 3 in Ref.~\cite{Wu:2023zmh}  and the mass values of the five channels in table~\ref{tab:mass} , the difference of $\Delta m = m_H - 2m_\Lambda$ can be estimated.  The results are:
\begin{align}
\Delta m = m_{H_1}  - 2m_\Lambda = -0.0026(30),\\
\Delta m = m_{H_{27}}  - 2m_\Lambda = 0.032(3)),\\
\Delta m = m_{H_{\Lambda\Lambda}}  - 2m_\Lambda = 0.009(3),\\
\Delta m = m_{H_{N\Xi}}  - 2m_\Lambda = -0.004(3),\\
\Delta m = m_{H_{\Sigma\Sigma}}  - 2m_\Lambda = -0.006(3).
\end{align}
Using the lattice spacing presented in table~\ref{tab:lattice_spacings}, the physical units can be obtained. From the results of $\Delta m$, one can find that
H-dibaryon realized through $H_1, H_{N\Xi}$ and $H_{\Sigma\Sigma}$ channels is stable  under strong interaction, while H-dibaryon realized through $H_{27}, H_{\Lambda\Lambda}$  channels 
are unstable. Other discussions concerning the binding energy of H-dibaryon can be found in Ref.~\cite{Wu:2023zmh} and therein.

In our simulation, the change of temperature is represented by the change of $T/T_c$. $T_c$ is pseudocritical temperature determined via renormalized
Polyakov loop and estimated to be $T_c = 185(4)\, {\rm MeV}$~\cite{Aarts:2014nba,Aarts:2020vyb}.

At different temperatures, the spectral density distribution of the five channels $H_1$, $H_{27}$, $H_{\Lambda\Lambda}$, $H_{N\Xi}$ and $H_{\Sigma\Sigma}$
are almost the same.  

At the lowest temperature $T/T_c = 0.24$, the spectral density distribution of the five channels has rich peak structure.
The mass spectrum of the five channels approximately reflects the peak position of spectral density distribution.
At the highest temperature, the spectrum density remains one peak structure indicating one particle state. Considering the high temperature, the one-peak structure at high temperature demands further investigation. More discussion about the spectral density can be found in Ref.~\cite{Wu:2023zmh}.

Our simualtions are at $m_\pi = 384(4) \ {\rm MeV}$ which is far from physical pion mass, so simulations with lower pion mass are expected to
give us more information about the properties of H-dibaryon.

\appendix

\begin{widetext}

\section{Correlator expressions}\label{appendix:a}
The correlation function $G_1$ for $H_1$  channel can be found in Ref.~\cite{Wetzorke:2001tgi}. The correlation functions 
$G_{27}$, $G_{\Lambda\Lambda}$,  $G_{N\Xi}$ and $G_{\Sigma\Sigma}$ of $H_{27}$, $H_{\Lambda\Lambda}$, $H_{N\Xi}$ and $H_{\Sigma\Sigma}$
are given by the following expressions:
\begin{align}
 G_{27} &= \epsilon_{abc}\epsilon_{def}\epsilon_{ghi}\epsilon_{jkl} (C\gamma_{5}P_+)_{\beta\gamma} (C\gamma_{5}P_+)_{\epsilon\phi} 
 (C\gamma_{5}P_+)_{\delta\alpha} (C\gamma_{5}P_+)_{\rho\sigma} (C\gamma_{5}P_+)_{\mu\nu}  (C\gamma_{5}P_+)_{\lambda\tau} \times \big\{ \nonumber \\  
    & 9 \times ( S_{\alpha a,\tau j} S_{\delta d,\lambda g} - S_{\alpha a,\lambda g} S_{\delta d,\tau j} )
( U_{\beta b,\nu k} U_{\epsilon e,\sigma h} - U_{\beta b,\sigma h} U_{\epsilon e,\nu k} )
( D_{\gamma c,\mu l} D_{\phi f,\rho i} - D_{\gamma c,\rho i} D_{\phi f,\mu l} )
 +  \nonumber \\
& 3 \times ( S_{\alpha a,\rho i} S_{\delta d,\lambda g} - S_{\alpha a,\lambda g} S_{\delta d,\rho i} )
( U_{\beta b,\tau j} U_{\epsilon e,\mu l} - U_{\beta b,\mu l} U_{\epsilon e,\tau j} )
( D_{\gamma c,\nu k} D_{\phi f,\sigma h} - D_{\gamma c,\sigma h} D_{\phi f,\nu k} )
 +  \nonumber \\
& 3 \times ( S_{\alpha a,\rho i} S_{\delta d,\lambda g} - S_{\alpha a,\lambda g} S_{\delta d,\rho i} )
( U_{\beta b,\nu k} U_{\epsilon e,\sigma h} - U_{\beta b,\sigma h} U_{\epsilon e,\nu k} )
( D_{\gamma c,\tau j} D_{\phi f,\mu l} - D_{\gamma c,\mu l} D_{\phi f,\tau j} )
 +  \nonumber \\
& 3 \times ( S_{\delta d,\lambda g} S_{\phi f,\tau j} - S_{\delta d,\tau j} S_{\phi f,\lambda g} )
( U_{\alpha a,\nu k} U_{\gamma c,\sigma h} - U_{\alpha a,\sigma h} U_{\gamma c,\nu k} )
( D_{\beta b,\mu l} D_{\epsilon e,\rho i} - D_{\beta b,\rho i} D_{\epsilon e,\mu l} )
 +  \nonumber \\
& ( S_{\delta d,\lambda g} S_{\phi f,\rho i} - S_{\delta d,\rho i} S_{\phi f,\lambda g} )
( U_{\alpha a,\tau j} U_{\gamma c,\mu l} - U_{\alpha a,\mu l} U_{\gamma c,\tau j} )
( D_{\beta b,\nu k} D_{\epsilon e,\sigma h} - D_{\beta b,\sigma h} D_{\epsilon e,\nu k} )
 +  \nonumber \\
& ( S_{\delta d,\lambda g} S_{\phi f,\rho i} - S_{\delta d,\rho i} S_{\phi f,\lambda g} )
( U_{\alpha a,\nu k} U_{\gamma c,\sigma h} - U_{\alpha a,\sigma h} U_{\gamma c,\nu k} )
( D_{\beta b,\tau j} D_{\epsilon e,\mu l} - D_{\beta b,\mu l} D_{\epsilon e,\tau j} )
 +  \nonumber \\
& 3 \times ( S_{\delta d,\lambda g} S_{\phi f,\tau j} - S_{\delta d,\tau j} S_{\phi f,\lambda g} )
( U_{\beta b,\nu k} U_{\epsilon e,\sigma h} - U_{\beta b,\sigma h} U_{\epsilon e,\nu k} )
( D_{\alpha a,\mu l} D_{\gamma c,\rho i} - D_{\alpha a,\rho i} D_{\gamma c,\mu l} )
 +  \nonumber \\
& ( S_{\delta d,\lambda g} S_{\phi f,\rho i} - S_{\delta d,\rho i} S_{\phi f,\lambda g} )
( U_{\beta b,\tau j} U_{\epsilon e,\mu l} - U_{\beta b,\mu l} U_{\epsilon e,\tau j} )
( D_{\alpha a,\nu k} D_{\gamma c,\sigma h} - D_{\alpha a,\sigma h} D_{\gamma c,\nu k} )
 +  \nonumber \\
& ( S_{\delta d,\lambda g} S_{\phi f,\rho i} - S_{\delta d,\rho i} S_{\phi f,\lambda g} )
( U_{\beta b,\nu k} U_{\epsilon e,\sigma h} - U_{\beta b,\sigma h} U_{\epsilon e,\nu k} )
( D_{\alpha a,\tau j} D_{\gamma c,\mu l} - D_{\alpha a,\mu l} D_{\gamma c,\tau j} ) \big\},
\end{align}

\begin{align}
 G_{\Lambda\Lambda} &= \epsilon_{abc}\epsilon_{def}\epsilon_{ghi}\epsilon_{jkl} (C\gamma_{5}P_+)_{\beta\gamma} (C\gamma_{5}P_+)_{\epsilon\phi}
 (C\gamma_{5}P_+)_{\delta\alpha} (C\gamma_{5}P_+)_{\rho\sigma} (C\gamma_{5}P_+)_{\mu\nu}  (C\gamma_{5}P_+)_{\lambda\tau} \times \big\{ \nonumber \\
     & ( S_{\alpha a,\tau j} S_{\delta d,\lambda g} - S_{\alpha a,\lambda g} S_{\delta d,\tau j} )
( U_{\beta b,\nu k} U_{\epsilon e,\sigma h} - U_{\beta b,\sigma h} U_{\epsilon e,\nu k} )
( D_{\gamma c,\mu l} D_{\phi f,\rho i} - D_{\gamma c,\rho i} D_{\phi f,\mu l} ) \big\},
\end{align}

\begin{align}
G_{N\Xi} &= \epsilon_{abc}\epsilon_{def}\epsilon_{ghi}\epsilon_{jkl} (C\gamma_{5}P_+)_{\beta\gamma} (C\gamma_{5}P_+)_{\epsilon\phi}
 (C\gamma_{5}P_+)_{\delta\alpha} (C\gamma_{5}P_+)_{\rho\sigma} (C\gamma_{5}P_+)_{\mu\nu}  (C\gamma_{5}P_+)_{\lambda\tau} \times \big\{ \nonumber \\
  & ( S_{\delta d,\lambda g} S_{\epsilon e,\sigma h} - S_{\delta d,\sigma h} S_{\epsilon e,\lambda g} )
( U_{\alpha a,\tau j} U_{\beta b,\nu k} - U_{\alpha a,\nu k} U_{\beta b,\tau j} )
( D_{\gamma c,\mu l} D_{\phi f,\rho i} - D_{\gamma c,\rho i} D_{\phi f,\mu l} )
 +  \nonumber \\
& ( S_{\delta d,\lambda g} S_{\epsilon e,\sigma h} - S_{\delta d,\sigma h} S_{\epsilon e,\lambda g} )
( U_{\alpha a,\rho i} U_{\beta b,\nu k} - U_{\alpha a,\nu k} U_{\beta b,\rho i} )
( D_{\gamma c,\mu l} D_{\phi f,\tau j} - D_{\gamma c,\tau j} D_{\phi f,\mu l} )
 +  \nonumber \\
& ( S_{\delta d,\tau j} S_{\epsilon e,\nu k} - S_{\delta d,\nu k} S_{\epsilon e,\tau j} )
( U_{\alpha a,\lambda g} U_{\beta b,\sigma h} - U_{\alpha a,\sigma h} U_{\beta b,\lambda g} )
( D_{\gamma c,\mu l} D_{\phi f,\rho i} - D_{\gamma c,\rho i} D_{\phi f,\mu l} )
 +  \nonumber \\
& ( S_{\delta d,\tau j} S_{\epsilon e,\nu k} - S_{\delta d,\nu k} S_{\epsilon e,\tau j} )
( U_{\alpha a,\sigma h} U_{\beta b,\mu l} - U_{\alpha a,\mu l} U_{\beta b,\sigma h} )
( D_{\gamma c,\rho i} D_{\phi f,\lambda g} - D_{\gamma c,\lambda g} D_{\phi f,\rho i} )
 +  \nonumber \\
& ( S_{\delta d,\lambda g} S_{\epsilon e,\sigma h} - S_{\delta d,\sigma h} S_{\epsilon e,\lambda g} )
( U_{\beta b,\nu k} U_{\phi f,\tau j} - U_{\beta b,\tau j} U_{\phi f,\nu k} )
( D_{\alpha a,\rho i} D_{\gamma c,\mu l} - D_{\alpha a,\mu l} D_{\gamma c,\rho i} )
 +  \nonumber \\
& ( S_{\delta d,\lambda g} S_{\epsilon e,\sigma h} - S_{\delta d,\sigma h} S_{\epsilon e,\lambda g} )
( U_{\beta b,\nu k} U_{\phi f,\rho i} - U_{\beta b,\rho i} U_{\phi f,\nu k} )
( D_{\alpha a,\tau j} D_{\gamma c,\mu l} - D_{\alpha a,\mu l} D_{\gamma c,\tau j} )
 +  \nonumber \\
& ( S_{\delta d,\tau j} S_{\epsilon e,\nu k} - S_{\delta d,\nu k} S_{\epsilon e,\tau j} )
( U_{\beta b,\sigma h} U_{\phi f,\lambda g} - U_{\beta b,\lambda g} U_{\phi f,\sigma h} )
( D_{\alpha a,\rho i} D_{\gamma c,\mu l} - D_{\alpha a,\mu l} D_{\gamma c,\rho i} )
 +  \nonumber \\
& ( S_{\delta d,\tau j} S_{\epsilon e,\nu k} - S_{\delta d,\nu k} S_{\epsilon e,\tau j} )
( U_{\beta b,\mu l} U_{\phi f,\sigma h} - U_{\beta b,\sigma h} U_{\phi f,\mu l} )
( D_{\alpha a,\lambda g} D_{\gamma c,\rho i} - D_{\alpha a,\rho i} D_{\gamma c,\lambda g} )
 +  \nonumber \\
& ( S_{\alpha a,\lambda g} S_{\beta b,\sigma h} - S_{\alpha a,\sigma h} S_{\beta b,\lambda g} )
( U_{\delta d,\tau j} U_{\epsilon e,\nu k} - U_{\delta d,\nu k} U_{\epsilon e,\tau j} )
( D_{\gamma c,\mu l} D_{\phi f,\rho i} - D_{\gamma c,\rho i} D_{\phi f,\mu l} )
 +  \nonumber \\
& ( S_{\alpha a,\lambda g} S_{\beta b,\sigma h} - S_{\alpha a,\sigma h} S_{\beta b,\lambda g} )
( U_{\delta d,\rho i} U_{\epsilon e,\nu k} - U_{\delta d,\nu k} U_{\epsilon e,\rho i} )
( D_{\gamma c,\mu l} D_{\phi f,\tau j} - D_{\gamma c,\tau j} D_{\phi f,\mu l} )
 +  \nonumber \\
& ( S_{\alpha a,\tau j} S_{\beta b,\nu k} - S_{\alpha a,\nu k} S_{\beta b,\tau j} )
( U_{\delta d,\lambda g} U_{\epsilon e,\sigma h} - U_{\delta d,\sigma h} U_{\epsilon e,\lambda g} )
( D_{\gamma c,\mu l} D_{\phi f,\rho i} - D_{\gamma c,\rho i} D_{\phi f,\mu l} )
 +  \nonumber \\
& ( S_{\alpha a,\tau j} S_{\beta b,\nu k} - S_{\alpha a,\nu k} S_{\beta b,\tau j} )
( U_{\delta d,\sigma h} U_{\epsilon e,\mu l} - U_{\delta d,\mu l} U_{\epsilon e,\sigma h} )
( D_{\gamma c,\rho i} D_{\phi f,\lambda g} - D_{\gamma c,\lambda g} D_{\phi f,\rho i} )
 +  \nonumber \\
  & ( S_{\alpha a,\lambda g} S_{\beta b,\sigma h} - S_{\alpha a,\sigma h} S_{\beta b,\lambda g} )
  ( U_{\gamma c,\nu k} U_{\epsilon e,\tau j} - U_{\gamma c,\tau j} U_{\epsilon e,\nu k} )
 ( D_{\delta d,\rho i} D_{\phi f,\mu l} - D_{\delta d,\mu l} D_{\phi f,\rho i} )
   +  \nonumber \\
 & ( S_{\alpha a,\lambda g} S_{\beta b,\sigma h} - S_{\alpha a,\sigma h} S_{\beta b,\lambda g} )
 ( U_{\gamma c,\nu k} U_{\epsilon e,\rho i} - U_{\gamma c,\rho i} U_{\epsilon e,\nu k} )
 ( D_{\delta d,\tau j} D_{\phi f,\mu l} - D_{\delta d,\mu l} D_{\phi f,\tau j} )
   +  \nonumber \\
 & ( S_{\alpha a,\tau j} S_{\beta b,\nu k} - S_{\alpha a,\nu k} S_{\beta b,\tau j} )
  ( U_{\gamma c,\sigma h} U_{\epsilon e,\lambda g} - U_{\gamma c,\lambda g} U_{\epsilon e,\sigma h} )
  ( D_{\delta d,\rho i} D_{\phi f,\mu l} - D_{\delta d,\mu l} D_{\phi f,\rho i} )
  +  \nonumber \\
  & ( S_{\alpha a,\tau j} S_{\beta b,\nu k} - S_{\alpha a,\nu k} S_{\beta b,\tau j} )
  ( U_{\gamma c,\mu l} U_{\epsilon e,\sigma h} - U_{\gamma c,\sigma h} U_{\epsilon e,\mu l} )
  ( D_{\delta d,\lambda g} D_{\phi f,\rho i} - D_{\delta d,\rho i} D_{\phi f,\lambda g} ) \big\},
  \end{align}

\begin{align}
G_{\Sigma\Sigma} &= \epsilon_{abc}\epsilon_{def}\epsilon_{ghi}\epsilon_{jkl} (C\gamma_{5}P_+)_{\beta\gamma} (C\gamma_{5}P_+)_{\epsilon\phi}
 (C\gamma_{5}P_+)_{\delta\alpha} (C\gamma_{5}P_+)_{\rho\sigma} (C\gamma_{5}P_+)_{\mu\nu}  (C\gamma_{5}P_+)_{\lambda\tau} \times \big\{ \nonumber \\
 & 4 \times ( S_{\gamma c,\mu l} S_{\phi f,\rho i} - S_{\gamma c,\rho i} S_{\phi f,\mu l} )
( U_{\alpha a,\tau j} U_{\beta b,\nu k} - U_{\alpha a,\nu k} U_{\beta b,\tau j} )
( D_{\delta d,\lambda g} D_{\epsilon e,\sigma h} - D_{\delta d,\sigma h} D_{\epsilon e,\lambda g} )
 +  \nonumber \\
& 2 \times ( S_{\gamma c,\mu l} S_{\phi f,\rho i} - S_{\gamma c,\rho i} S_{\phi f,\mu l} )
( U_{\alpha a,\nu k} U_{\beta b,\lambda g} - U_{\alpha a,\lambda g} U_{\beta b,\nu k} )
( D_{\delta d,\sigma h} D_{\epsilon e,\tau j} - D_{\delta d,\tau j} D_{\epsilon e,\sigma h} )
 +  \nonumber \\
& 2 \times ( S_{\gamma c,\mu l} S_{\phi f,\rho i} - S_{\gamma c,\rho i} S_{\phi f,\mu l} )
( U_{\alpha a,\sigma h} U_{\beta b,\nu k} - U_{\alpha a,\nu k} U_{\beta b,\sigma h} )
( D_{\delta d,\lambda g} D_{\epsilon e,\tau j} - D_{\delta d,\tau j} D_{\epsilon e,\lambda g} )
 +  \nonumber \\
& 2 \times ( S_{\gamma c,\mu l} S_{\phi f,\rho i} - S_{\gamma c,\rho i} S_{\phi f,\mu l} )
( U_{\alpha a,\tau j} U_{\beta b,\sigma h} - U_{\alpha a,\sigma h} U_{\beta b,\tau j} )
( D_{\delta d,\lambda g} D_{\epsilon e,\nu k} - D_{\delta d,\nu k} D_{\epsilon e,\lambda g} )
 +  \nonumber \\
& 2 \times ( S_{\gamma c,\mu l} S_{\phi f,\rho i} - S_{\gamma c,\rho i} S_{\phi f,\mu l} )
( U_{\alpha a,\lambda g} U_{\beta b,\tau j} - U_{\alpha a,\tau j} U_{\beta b,\lambda g} )
( D_{\delta d,\sigma h} D_{\epsilon e,\nu k} - D_{\delta d,\nu k} D_{\epsilon e,\sigma h} )
 +  \nonumber \\
& 4 \times ( S_{\gamma c,\mu l} S_{\phi f,\rho i} - S_{\gamma c,\rho i} S_{\phi f,\mu l} )
( U_{\alpha a,\lambda g} U_{\beta b,\sigma h} - U_{\alpha a,\sigma h} U_{\beta b,\lambda g} )
( D_{\delta d,\tau j} D_{\epsilon e,\nu k} - D_{\delta d,\nu k} D_{\epsilon e,\tau j} )
 +  \nonumber \\
& 2 \times ( S_{\gamma c,\mu l} S_{\phi f,\rho i} - S_{\gamma c,\rho i} S_{\phi f,\mu l} )
( U_{\beta b,\tau j} U_{\delta d,\nu k} - U_{\beta b,\nu k} U_{\delta d,\tau j} )
( D_{\alpha a,\sigma h} D_{\epsilon e,\lambda g} - D_{\alpha a,\lambda g} D_{\epsilon e,\sigma h} )
 +  \nonumber \\
& ( S_{\gamma c,\mu l} S_{\phi f,\rho i} - S_{\gamma c,\rho i} S_{\phi f,\mu l} )
( U_{\beta b,\nu k} U_{\delta d,\lambda g} - U_{\beta b,\lambda g} U_{\delta d,\nu k} )
( D_{\alpha a,\tau j} D_{\epsilon e,\sigma h} - D_{\alpha a,\sigma h} D_{\epsilon e,\tau j} )
 +  \nonumber \\
& ( S_{\gamma c,\mu l} S_{\phi f,\rho i} - S_{\gamma c,\rho i} S_{\phi f,\mu l} )
( U_{\beta b,\sigma h} U_{\delta d,\nu k} - U_{\beta b,\nu k} U_{\delta d,\sigma h} )
( D_{\alpha a,\tau j} D_{\epsilon e,\lambda g} - D_{\alpha a,\lambda g} D_{\epsilon e,\tau j} )
 +  \nonumber \\
& ( S_{\gamma c,\mu l} S_{\phi f,\rho i} - S_{\gamma c,\rho i} S_{\phi f,\mu l} )
( U_{\beta b,\tau j} U_{\delta d,\sigma h} - U_{\beta b,\sigma h} U_{\delta d,\tau j} )
( D_{\alpha a,\nu k} D_{\epsilon e,\lambda g} - D_{\alpha a,\lambda g} D_{\epsilon e,\nu k} )
 +  \nonumber \\
& ( S_{\gamma c,\mu l} S_{\phi f,\rho i} - S_{\gamma c,\rho i} S_{\phi f,\mu l} )
( U_{\beta b,\lambda g} U_{\delta d,\tau j} - U_{\beta b,\tau j} U_{\delta d,\lambda g} )
( D_{\alpha a,\nu k} D_{\epsilon e,\sigma h} - D_{\alpha a,\sigma h} D_{\epsilon e,\nu k} )
 +  \nonumber \\
& 2 \times ( S_{\gamma c,\mu l} S_{\phi f,\rho i} - S_{\gamma c,\rho i} S_{\phi f,\mu l} )
( U_{\beta b,\lambda g} U_{\delta d,\sigma h} - U_{\beta b,\sigma h} U_{\delta d,\lambda g} )
( D_{\alpha a,\nu k} D_{\epsilon e,\tau j} - D_{\alpha a,\tau j} D_{\epsilon e,\nu k} )
 +  \nonumber \\
& 2 \times ( S_{\gamma c,\mu l} S_{\phi f,\rho i} - S_{\gamma c,\rho i} S_{\phi f,\mu l} )
( U_{\beta b,\nu k} U_{\epsilon e,\tau j} - U_{\beta b,\tau j} U_{\epsilon e,\nu k} )
( D_{\alpha a,\sigma h} D_{\delta d,\lambda g} - D_{\alpha a,\lambda g} D_{\delta d,\sigma h} )
 +  \nonumber \\
& ( S_{\gamma c,\mu l} S_{\phi f,\rho i} - S_{\gamma c,\rho i} S_{\phi f,\mu l} )
( U_{\beta b,\lambda g} U_{\epsilon e,\nu k} - U_{\beta b,\nu k} U_{\epsilon e,\lambda g} )
( D_{\alpha a,\tau j} D_{\delta d,\sigma h} - D_{\alpha a,\sigma h} D_{\delta d,\tau j} )
 +  \nonumber \\
& ( S_{\gamma c,\mu l} S_{\phi f,\rho i} - S_{\gamma c,\rho i} S_{\phi f,\mu l} )
( U_{\beta b,\nu k} U_{\epsilon e,\sigma h} - U_{\beta b,\sigma h} U_{\epsilon e,\nu k} )
( D_{\alpha a,\tau j} D_{\delta d,\lambda g} - D_{\alpha a,\lambda g} D_{\delta d,\tau j} )
 +  \nonumber \\
& ( S_{\gamma c,\mu l} S_{\phi f,\rho i} - S_{\gamma c,\rho i} S_{\phi f,\mu l} )
( U_{\beta b,\sigma h} U_{\epsilon e,\tau j} - U_{\beta b,\tau j} U_{\epsilon e,\sigma h} )
( D_{\alpha a,\nu k} D_{\delta d,\lambda g} - D_{\alpha a,\lambda g} D_{\delta d,\nu k} )
 +  \nonumber \\
& ( S_{\gamma c,\mu l} S_{\phi f,\rho i} - S_{\gamma c,\rho i} S_{\phi f,\mu l} )
( U_{\beta b,\tau j} U_{\epsilon e,\lambda g} - U_{\beta b,\lambda g} U_{\epsilon e,\tau j} )
( D_{\alpha a,\nu k} D_{\delta d,\sigma h} - D_{\alpha a,\sigma h} D_{\delta d,\nu k} )
 +  \nonumber \\
& 2 \times ( S_{\gamma c,\mu l} S_{\phi f,\rho i} - S_{\gamma c,\rho i} S_{\phi f,\mu l} )
( U_{\beta b,\sigma h} U_{\epsilon e,\lambda g} - U_{\beta b,\lambda g} U_{\epsilon e,\sigma h} )
( D_{\alpha a,\nu k} D_{\delta d,\tau j} - D_{\alpha a,\tau j} D_{\delta d,\nu k} )
 +  \nonumber \\
& 2 \times ( S_{\gamma c,\mu l} S_{\phi f,\rho i} - S_{\gamma c,\rho i} S_{\phi f,\mu l} )
( U_{\alpha a,\tau j} U_{\epsilon e,\nu k} - U_{\alpha a,\nu k} U_{\epsilon e,\tau j} )
( D_{\beta b,\sigma h} D_{\delta d,\lambda g} - D_{\beta b,\lambda g} D_{\delta d,\sigma h} )
 +  \nonumber \\
& ( S_{\gamma c,\mu l} S_{\phi f,\rho i} - S_{\gamma c,\rho i} S_{\phi f,\mu l} )
( U_{\alpha a,\nu k} U_{\epsilon e,\lambda g} - U_{\alpha a,\lambda g} U_{\epsilon e,\nu k} )
( D_{\beta b,\tau j} D_{\delta d,\sigma h} - D_{\beta b,\sigma h} D_{\delta d,\tau j} )
 +  \nonumber \\
& ( S_{\gamma c,\mu l} S_{\phi f,\rho i} - S_{\gamma c,\rho i} S_{\phi f,\mu l} )
( U_{\alpha a,\sigma h} U_{\epsilon e,\nu k} - U_{\alpha a,\nu k} U_{\epsilon e,\sigma h} )
( D_{\beta b,\tau j} D_{\delta d,\lambda g} - D_{\beta b,\lambda g} D_{\delta d,\tau j} )
 +  \nonumber \\
& ( S_{\gamma c,\mu l} S_{\phi f,\rho i} - S_{\gamma c,\rho i} S_{\phi f,\mu l} )
( U_{\alpha a,\tau j} U_{\epsilon e,\sigma h} - U_{\alpha a,\sigma h} U_{\epsilon e,\tau j} )
( D_{\beta b,\nu k} D_{\delta d,\lambda g} - D_{\beta b,\lambda g} D_{\delta d,\nu k} )
 +  \nonumber \\
& ( S_{\gamma c,\mu l} S_{\phi f,\rho i} - S_{\gamma c,\rho i} S_{\phi f,\mu l} )
( U_{\alpha a,\lambda g} U_{\epsilon e,\tau j} - U_{\alpha a,\tau j} U_{\epsilon e,\lambda g} )
( D_{\beta b,\nu k} D_{\delta d,\sigma h} - D_{\beta b,\sigma h} D_{\delta d,\nu k} )
 +  \nonumber \\
& 2 \times ( S_{\gamma c,\mu l} S_{\phi f,\rho i} - S_{\gamma c,\rho i} S_{\phi f,\mu l} )
( U_{\alpha a,\lambda g} U_{\epsilon e,\sigma h} - U_{\alpha a,\sigma h} U_{\epsilon e,\lambda g} )
( D_{\beta b,\nu k} D_{\delta d,\tau j} - D_{\beta b,\tau j} D_{\delta d,\nu k} )
 +  \nonumber \\
& 2 \times ( S_{\gamma c,\mu l} S_{\phi f,\rho i} - S_{\gamma c,\rho i} S_{\phi f,\mu l} )
( U_{\alpha a,\nu k} U_{\delta d,\tau j} - U_{\alpha a,\tau j} U_{\delta d,\nu k} )
( D_{\beta b,\sigma h} D_{\epsilon e,\lambda g} - D_{\beta b,\lambda g} D_{\epsilon e,\sigma h} )
 +  \nonumber \\
& ( S_{\gamma c,\mu l} S_{\phi f,\rho i} - S_{\gamma c,\rho i} S_{\phi f,\mu l} )
( U_{\alpha a,\lambda g} U_{\delta d,\nu k} - U_{\alpha a,\nu k} U_{\delta d,\lambda g} )
( D_{\beta b,\tau j} D_{\epsilon e,\sigma h} - D_{\beta b,\sigma h} D_{\epsilon e,\tau j} )
 +  \nonumber \\
& ( S_{\gamma c,\mu l} S_{\phi f,\rho i} - S_{\gamma c,\rho i} S_{\phi f,\mu l} )
( U_{\alpha a,\nu k} U_{\delta d,\sigma h} - U_{\alpha a,\sigma h} U_{\delta d,\nu k} )
( D_{\beta b,\tau j} D_{\epsilon e,\lambda g} - D_{\beta b,\lambda g} D_{\epsilon e,\tau j} )
 +  \nonumber \\
& ( S_{\gamma c,\mu l} S_{\phi f,\rho i} - S_{\gamma c,\rho i} S_{\phi f,\mu l} )
( U_{\alpha a,\sigma h} U_{\delta d,\tau j} - U_{\alpha a,\tau j} U_{\delta d,\sigma h} )
( D_{\beta b,\nu k} D_{\epsilon e,\lambda g} - D_{\beta b,\lambda g} D_{\epsilon e,\nu k} )
 +  \nonumber \\
& ( S_{\gamma c,\mu l} S_{\phi f,\rho i} - S_{\gamma c,\rho i} S_{\phi f,\mu l} )
( U_{\alpha a,\tau j} U_{\delta d,\lambda g} - U_{\alpha a,\lambda g} U_{\delta d,\tau j} )
( D_{\beta b,\nu k} D_{\epsilon e,\sigma h} - D_{\beta b,\sigma h} D_{\epsilon e,\nu k} )
 +  \nonumber \\
& 2 \times ( S_{\gamma c,\mu l} S_{\phi f,\rho i} - S_{\gamma c,\rho i} S_{\phi f,\mu l} )
( U_{\alpha a,\sigma h} U_{\delta d,\lambda g} - U_{\alpha a,\lambda g} U_{\delta d,\sigma h} )
( D_{\beta b,\nu k} D_{\epsilon e,\tau j} - D_{\beta b,\tau j} D_{\epsilon e,\nu k} )
 +  \nonumber \\
& 4 \times ( S_{\gamma c,\mu l} S_{\phi f,\rho i} - S_{\gamma c,\rho i} S_{\phi f,\mu l} )
( U_{\delta d,\tau j} U_{\epsilon e,\nu k} - U_{\delta d,\nu k} U_{\epsilon e,\tau j} )
( D_{\alpha a,\lambda g} D_{\beta b,\sigma h} - D_{\alpha a,\sigma h} D_{\beta b,\lambda g} )
 +  \nonumber \\
& 2 \times ( S_{\gamma c,\mu l} S_{\phi f,\rho i} - S_{\gamma c,\rho i} S_{\phi f,\mu l} )
( U_{\delta d,\nu k} U_{\epsilon e,\lambda g} - U_{\delta d,\lambda g} U_{\epsilon e,\nu k} )
( D_{\alpha a,\sigma h} D_{\beta b,\tau j} - D_{\alpha a,\tau j} D_{\beta b,\sigma h} )
 +  \nonumber \\
& 2 \times ( S_{\gamma c,\mu l} S_{\phi f,\rho i} - S_{\gamma c,\rho i} S_{\phi f,\mu l} )
( U_{\delta d,\sigma h} U_{\epsilon e,\nu k} - U_{\delta d,\nu k} U_{\epsilon e,\sigma h} )
( D_{\alpha a,\lambda g} D_{\beta b,\tau j} - D_{\alpha a,\tau j} D_{\beta b,\lambda g} )
 +  \nonumber \\
& 2 \times ( S_{\gamma c,\mu l} S_{\phi f,\rho i} - S_{\gamma c,\rho i} S_{\phi f,\mu l} )
( U_{\delta d,\tau j} U_{\epsilon e,\sigma h} - U_{\delta d,\sigma h} U_{\epsilon e,\tau j} )
( D_{\alpha a,\lambda g} D_{\beta b,\nu k} - D_{\alpha a,\nu k} D_{\beta b,\lambda g} )
 +  \nonumber \\
& 2 \times ( S_{\gamma c,\mu l} S_{\phi f,\rho i} - S_{\gamma c,\rho i} S_{\phi f,\mu l} )
( U_{\delta d,\lambda g} U_{\epsilon e,\tau j} - U_{\delta d,\tau j} U_{\epsilon e,\lambda g} )
( D_{\alpha a,\sigma h} D_{\beta b,\nu k} - D_{\alpha a,\nu k} D_{\beta b,\sigma h} )
 +  \nonumber \\
& 4 \times ( S_{\gamma c,\mu l} S_{\phi f,\rho i} - S_{\gamma c,\rho i} S_{\phi f,\mu l} )
( U_{\delta d,\lambda g} U_{\epsilon e,\sigma h} - U_{\delta d,\sigma h} U_{\epsilon e,\lambda g} )
( D_{\alpha a,\tau j} D_{\beta b,\nu k} - D_{\alpha a,\nu k} D_{\beta b,\tau j} ) \big\}.
\end{align}

\end{widetext}

\begin{acknowledgments}
We thank Gert Aarts, Simon Hands, Chris Allton, and Jonas Glesaaen  for valuable helps.
 We modify
 the adapted version of  OpenQCD code~\cite{openqcd} and MILC code~\cite{MILC} to carry out this simulation and we use the computer program~\cite{Hansen:code} to
 calculate the spectral density of correlation function.
 The adaptation of  OpenQCD code is  publicly available~\cite{fastsum1}. 
 The simulations are carried out on $N_f=2+1$ Generation2 (Gen2)
FASTSUM ensembles~\cite{Aarts:2020vyb} of which the ensembles at the lowest temperature are provided by the HadSpec
collaboration~\cite{Edwards:2008ja,HadronSpectrum:2008xlg}.
  This work was supported by the National Natural Science Foundation of China (NSFC) (11347029). This study was conducted at the high performance computing platform of Ji
angsu University.

\end{acknowledgments}


\begin{thebibliography}{99}



\bibitem{Jaffe:1976yi}
R.~L.~Jaffe,
Phys. Rev. Lett. \textbf{38}, 195-198 (1977)
[erratum: Phys. Rev. Lett. \textbf{38}, 617(E) (1977)]

\bibitem{Farrar:2017eqq}
G.~R.~Farrar,
arXiv:1708.08951 [hep-ph].


\bibitem{Aoki:1991ip}
S.~Aoki, S.~Y.~Bahk, K.~S.~Chung, S.~H.~Chung, H.~Funahashi, C.~H.~Hahn, T.~Hara, S.~Hirata, K.~Hoshino and M.~Ieiri, \textit{et al.}
Prog. Theor. Phys. \textbf{85}, 1287-1298 (1991).

\bibitem{KEK-PSE224:1998trj}
J.~K.~Ahn \textit{et al.} [KEK-PS E224],
Phys. Lett. B \textbf{444}, 267-272 (1998).

\bibitem{Takahashi:2001nm}
H.~Takahashi, J.~K.~Ahn, H.~Akikawa, S.~Aoki, K.~Arai, S.~Y.~Bahk, K.~M.~Baik, B.~Bassalleck, J.~H.~Chung and M.~S.~Chung, \textit{et al.}
Phys. Rev. Lett. \textbf{87}, 212502 (2001).

\bibitem{Yoon:2007aq}
C.~J.~Yoon, H.~Akikawa, K.~Aoki, Y.~Fukao, H.~Funahashi, M.~Hayata, K.~Imai, K.~Miwa, H.~Okada and N.~Saito, \textit{et al.}
Phys. Rev. C \textbf{75}, 022201(R) (2007).

\bibitem{KEKE176:2009jzw}
S.~Aoki \textit{et al.} [KEK E176],
Nucl. Phys. A \textbf{828}, 191-232 (2009).

\bibitem{Nakazawa:2010zza}
K.~Nakazawa [KEK-E176 and J-PARC-E07],
Nucl. Phys. A \textbf{835}, 207-214 (2010).

\bibitem{Ekawa:2018oqt}
H.~Ekawa, K.~Agari, J.~K.~Ahn, T.~Akaishi, Y.~Akazawa, S.~Ashikaga, B.~Bassalleck, S.~Bleser, Y.~Endo and Y.~Fujikawa, \textit{et al.}
PTEP \textbf{2019}, no.2, 021D02 (2019).


\bibitem{Belle:2013sba}
B.~H.~Kim \textit{et al.} [Belle],
Phys. Rev. Lett. \textbf{110}, no.22, 222002 (2013).

\bibitem{BaBar:2018hpv}
J.~P.~Lees \textit{et al.} [BaBar],
Phys. Rev. Lett. \textbf{122}, no.7, 072002 (2019).


\bibitem{Iwasaki:1987db}
Y.~Iwasaki, T.~Yoshie and Y.~Tsuboi,
Phys. Rev. Lett. \textbf{60}, 1371-1374 (1988).

\bibitem{Luo:2011ar}
Z.~H.~Luo, M.~Loan and Y.~Liu,
Phys. Rev. D \textbf{84}, 034502 (2011).

\bibitem{Luo:2007zzb}
Z.~H.~Luo, M.~Loan and X.~Q.~Luo,
Mod. Phys. Lett. A \textbf{22}, 591-597 (2007).

\bibitem{Mackenzie:1985vv}
P.~B.~Mackenzie and H.~B.~Thacker,
Phys. Rev. Lett. \textbf{55}, 2539 (1985).

\bibitem{Pochinsky:1998zi}
A.~Pochinsky, J.~W.~Negele and B.~Scarlet,
Nucl. Phys. B Proc. Suppl. \textbf{73}, 255-257 (1999).

\bibitem{Wetzorke:1999rt}
I.~Wetzorke, F.~Karsch and E.~Laermann,
Nucl. Phys. B Proc. Suppl. \textbf{83}, 218-220 (2000).

\bibitem{Wetzorke:2002mx}
I.~Wetzorke and F.~Karsch,
Nucl. Phys. B Proc. Suppl. \textbf{119}, 278-280 (2003).






\bibitem{Beane:2009py}
S.~R.~Beane, W.~Detmold, H.-W.~Lin, T.~C.~Luu, K.~Orginos, M.~J.~Savage, A.~Torok and A.~Walker-Loud [NPLQCD],
Phys. Rev. D \textbf{81}, 054505 (2010).
\bibitem{NPLQCD:2010ocs}
S.~R.~Beane \textit{et al.} [NPLQCD],
Phys. Rev. Lett. \textbf{106}, 162001 (2011).

\bibitem{Beane:2011zpa}
S.~R.~Beane, E.~Chang, W.~Detmold, B.~Joo, H.~W.~Lin, T.~C.~Luu, K.~Orginos, A.~Parreno, M.~J.~Savage and A.~Torok, \textit{et al.}
Mod. Phys. Lett. A \textbf{26}, 2587-2595 (2011).

\bibitem{NPLQCD:2011naw}
S.~R.~Beane \textit{et al.} [NPLQCD],
Phys. Rev. D \textbf{85}, 054511 (2012).

\bibitem{NPLQCD:2012mex}
S.~R.~Beane \textit{et al.} [NPLQCD],
Phys. Rev. D \textbf{87}, no.3, 034506 (2013).



\bibitem{Inoue:2010hs}
T.~Inoue \textit{et al.} [HAL QCD],
Prog. Theor. Phys. \textbf{124}, 591-603 (2010).

\bibitem{Inoue:2010es}
T.~Inoue,  N.~Ishii, S.~Aoki,  T.~Doi,  T.~Hatsuda,  Y.~Ikeda, K.~ Murano,   H.~Nemura,  and K.~Sasaki [HAL QCD],
Phys. Rev. Lett. \textbf{106}, 162002 (2011).

\bibitem{Inoue:2011ai}
T.~Inoue \textit{et al.} [HAL QCD],
Nucl. Phys. A \textbf{881}, 28-43 (2012).

\bibitem{HALQCD:2019wsz}
K.~Sasaki \textit{et al.} [HAL QCD],
Nucl. Phys. A \textbf{998}, 121737 (2020).

\bibitem{Sasaki:2016gpc}
K.~Sasaki, S.~Aoki, T.~Doi, S.~Gongyo, T.~Hatsuda, Y.~Ikeda, T.~Inoue, T.~Iritani, N.~Ishii and T.~Miyamoto, \textit{et al.}
PoS \textbf{LATTICE2015}, 088 (2016).

\bibitem{HALQCD:2018lur}
K.~Sasaki \textit{et al.} [HAL QCD],
EPJ Web Conf. \textbf{175}, 05010 (2018).
\bibitem{Francis:2018qch}
A.~Francis, J.~R.~Green, P.~M.~Junnarkar, C.~Miao, T.~D.~Rae and H.~Wittig,
Phys. Rev. D \textbf{99}, no.7, 074505 (2019).

\bibitem{Green:2021qol}
J.~R.~Green, A.~D.~Hanlon, P.~M.~Junnarkar and H.~Wittig,
Phys. Rev. Lett. \textbf{127}, no.24, 242003 (2021).

\bibitem{Junnarkar:2024kwd}
P.~M.~Junnarkar and N.~Mathur,
Phys. Rev. D \textbf{111}, no.1, 1 (2025)
doi:10.1103/PhysRevD.111.014512
[arXiv:2410.08519 [hep-lat]].

\bibitem{Luscher:1986pf}
M.~Luscher,
Commun. Math. Phys. \textbf{105}, 153-188 (1986).

\bibitem{Luscher:1990ux}
M.~Luscher,
Nucl. Phys. B \textbf{354}, 531-578 (1991).

\bibitem{Beane:2003da}
S.~R.~Beane, P.~F.~Bedaque, A.~Parreno and M.~J.~Savage,
Phys. Lett. B \textbf{585}, 106-114 (2004).

\bibitem{Beane:2006mx}
S.~R.~Beane, P.~F.~Bedaque, K.~Orginos and M.~J.~Savage,
Phys. Rev. Lett. \textbf{97}, 012001 (2006).


\bibitem{Lyu:2022tsd}
Y.~Lyu, H.~Tong, T.~Sugiura, S.~Aoki, T.~Doi, T.~Hatsuda, J.~Meng and T.~Miyamoto,
Phys. Rev. D \textbf{105}, no.7, 074512 (2022)
doi:10.1103/PhysRevD.105.074512
[arXiv:2201.02782 [hep-lat]].

\bibitem{Xing:2025uai}
H.~Xing, Y.~Geng, C.~Liu, L.~Liu, P.~Sun, J.~Wu, Z.~Yan and R.~Zhu,
Chin. Phys. C \textbf{49}, no.6, 063107 (2025)
doi:10.1088/1674-1137/adc11f
[arXiv:2502.05546 [hep-lat]].







\bibitem{Junnarkar:2022yak}
P.~M.~Junnarkar and N.~Mathur,
Phys. Rev. D \textbf{106}, no.5, 054511 (2022).
\bibitem{Mathur:2022ovu}
N.~Mathur, M.~Padmanath and D.~Chakraborty,
Phys. Rev. Lett. \textbf{130}, no.11, 111901 (2023).


\bibitem{Aarts:2010ek}
G.~Aarts, S.~Kim, M.~P.~Lombardo, M.~B.~Oktay, S.~M.~Ryan, D.~K.~Sinclair and J.~I.~Skullerud,
Phys. Rev. Lett. \textbf{106}, 061602 (2011).

\bibitem{Aarts:2011sm}
G.~Aarts, C.~Allton, S.~Kim, M.~P.~Lombardo, M.~B.~Oktay, S.~M.~Ryan, D.~K.~Sinclair and J.~I.~Skullerud,
JHEP \textbf{11}, 103 (2011).

\bibitem{Aarts:2013kaa}
G.~Aarts, C.~Allton, S.~Kim, M.~P.~Lombardo, S.~M.~Ryan and J.~I.~Skullerud,
JHEP \textbf{12}, 064 (2013).

\bibitem{Aarts:2014cda}
G.~Aarts, C.~Allton, T.~Harris, S.~Kim, M.~P.~Lombardo, S.~M.~Ryan and J.~I.~Skullerud,
JHEP \textbf{07}, 097 (2014).

\bibitem{Kelly:2018hsi}
A.~Kelly, A.~Rothkopf and J.~I.~Skullerud,
Phys. Rev. D \textbf{97}, no.11, 114509 (2018).

\bibitem{Aarts:2020vyb}
G.~Aarts, C.~Allton, J.~Glesaaen, S.~Hands, B.~J\"ager, S.~Kim, M.~P.~Lombardo, A.~A.~Nikolaev, S.~M.~Ryan and J.~I.~Skullerud, \textit{et al.}
Phys. Rev. D \textbf{105}, no.3, 034504 (2022).

\bibitem{Karsch:2003jg}
  F.~Karsch and E.~Laermann,
  Thermodynamics and in medium hadron properties from lattice QCD,
  Edited by Hwa, R.C. (ed.) et al.: Quark gluon plasma (World Scientific, Singapore, 2004) p. 1-59.

\bibitem{Aarts:2018glk}
G.~Aarts, C.~Allton, D.~De Boni and B.~J\"ager,
Phys. Rev. D \textbf{99}, no.7, 074503 (2019).

\bibitem{DeTar:1987ar}
  C.~E.~Detar and J.~B.~Kogut,
  Phys.\ Rev.\ Lett.\  {\bf 59} (1987) 399;
  Phys.\ Rev.\ D {\bf 36} (1987) 2828.

\bibitem{Pushkina:2004wa}
  I.~Pushkina {\it et al.} [QCD-TARO Collaboration],
  Phys.\ Lett.\ B {\bf 609} (2005) 265.

\bibitem{Datta:2012fz}
  S.~Datta, S.~Gupta, M.~Padmanath, J.~Maiti and N.~Mathur,
  JHEP {\bf 1302} (2013) 145.



\bibitem{Aarts:2015mma}
  G.~Aarts, C.~Allton, S.~Hands, B.~J\"ager, C.~Praki and J.~I.~Skullerud,
  Phys.\ Rev.\ D {\bf 92} (2015) no.1,  014503.


\bibitem{Wu:2023zmh}
L.~K.~Wu, H.~Tang, N.~Li and X.~Y.~Wang,
Chin. Phys. C \textbf{48}, no.8, 083105 (2024)
doi:10.1088/1674-1137/ad3d4c
[arXiv:2309.10258 [hep-lat]].


\bibitem{Donoghue:1986zd}
J.~Donoghue, E.~Golowich,  and B.~R. Holstein,
Phys. Rev. D \textbf{34},3434, (1986).

\bibitem{Golowich:1992zw}
E.~Golowich and T.~Sotirelis,
Phys. Rev. D \textbf{ 46},354, (1992).

\bibitem{Wetzorke:2001tgi}
I.~Wetzorke, https://pub.uni{-}bielefeld.de/record/2394939.

\bibitem{Hansen:2019idp}
M.~Hansen, A.~Lupo and N.~Tantalo,
Phys. Rev. D \textbf{99}, no.9, 094508 (2019).


\bibitem{Edwards:2008ja}
R.~G.~Edwards, B.~Joo and H.~W.~Lin,
Phys. Rev. D \textbf{78}, 054501 (2008).

\bibitem{HadronSpectrum:2008xlg}
H.~W.~Lin \textit{et al.} [Hadron Spectrum],
Phys. Rev. D \textbf{79}, 034502 (2009).



\bibitem{Luscher:2007es}
M.~Luscher,
JHEP \textbf{12}, 011 (2007).
\bibitem{Luscher:2007se}
M.~Luscher,
JHEP \textbf{07}, 081 (2007).

\bibitem{Morningstar:2003gk}
C.~Morningstar and M.~J.~Peardon,
Phys. Rev. D \textbf{69}, 054501 (2004).

\bibitem{Gusken:1989ad}
S.~Gusken, U.~Low, K.~H.~Mutter, R.~Sommer, A.~Patel and K.~Schilling,
Phys. Lett. B \textbf{227}, 266-269 (1989).

\bibitem{Aarts:2014nba}
G.~Aarts, C.~Allton, A.~Amato, P.~Giudice, S.~Hands and J.~I.~Skullerud,
JHEP \textbf{02}, 186 (2015).



\bibitem{openqcd}
 OpenQCD, luscher.web.cern.ch/luscher/openQCD/.

\bibitem{MILC}
https://web.physics.utah.edu/{\textasciitilde}detar/milc.html

\bibitem{Hansen:code}
 https://github.com/mrlhansen/rmsd.
 
 \bibitem{fastsum1}
 FASTSUM Collaboration, http://fastsum.gitlab.io/.

\end{thebibliography}
\end{document}